# Sensing-Aided Peer-to-Peer Millimeter-Wave Communication

Xiangyu Li[1,2], Sidong Guo[1] and Shez Malik[1]

[1] Georgia Institute of Technology, Atlanta 30332, USA
[2] Georgia Tech Shenzhen Institute, Tianjin University, Shenzhen 518055, China
xli985@gatech.edu

**Abstract.** One of the bottlenecks of modern communications is to enable sensing and mutual communication simultaneously without causing scheduling conflicts, and how sensing may be leveraged to help directional communication accuracy. To this end, we propose and implement a novel peer-to-peer (P2P) millimeter-wave communication system to jointly achieve beamforming and sensing. A radar and IMU-assisted tracking and beamforming algorithm is designed and tested. The results show that a robust tracking capacity with an overall higher throughput can be obtained. It is also hopeful that based on our proposed system, our design and implementation can be deployed in a more scalable manner for future extensions.

## 1 Introduction

Recent years have witnessed an increasing demand for high data rates in support of wireless communications. One of the most promising 5G and beyond wireless technologies is Millimeter-wave (mmWave) communications, which is motivated by the global spectrum shortage even up to present 5G frequencies [1]. It is attracting significant attention because of its huge amounts of spectrum, high propagation gains and other channel spatial features [2]. Next-generation communications will rely heavily on sensors, including radars, lidars, and cameras [3], for client-to-client communications with higher quality of service (QoS) and efficiency. By leveraging these sensing technologies before the data transmission stage, the real locations of clients in need of mutual communications can be obtained [4] and continuous detection and tracking is achievable in a real-time manner. These will make the best use of energy for transmitting data, especially when directionality is a required condition for effective mmWave beamforming [5], and largely avoid energy wastage and reach the goal of green and secure communications [6].

## 1.1 Related Work

Using mmWave for sensing-related purposes has been studied extensively. A case of mmWave band utilization for enhanced outdoor local area (eLA) access in 5G was investigated by Ghosh [7]. In the system, peak data rates of more than 10 Gbps and edge data rates in an excess of 100 Mbps could be achieved with huge bandwidth of mmWave. It was also demonstrated that in mmWave-supported cellular access networks, higher coverage and capacity are likely to be obtained [8][9]. With directional antennas, capacity gains, power, and spectral efficiency could be largely improved in the potential device-to-device (D2D) communications [10]. This has inspired us to further investigate the effects of mmWave directionality for the indoor scenario as a start point.

For indoor scenarios of mmWave communications, schemes with different settings were investigated to improve the QoS. In [1], the spatial modulation multiple-input-multiple-output (MIMO)-based scheme in indoor line-of-sight (LOS) mmWave communication at 60 GHz was studied. The error performance of spatial modulation was optimized in the theoretical aspect by maximizing the minimum Euclidean distance of received symbols. However, due to the directionality of mmWave, the location and orientation of target clients with sensing need to be further studied for higher accuracy. In [2], a radar sensing-assisted mmWave communication scheme was designed, but the results were obtained in a two-dimensional (2-D) environment via simulation while no field tests are conducted. The survey [4] introduced an approach for indoor detection and tracking of humans using mmWave sensor, where no mutual communications were involved for error correction. In addition, beamforming in mmWave communications were also deeply discussed based on its large bandwidth, unique channel features and hardware constraints in [11]. The precision of beamforming direction is suggested to improve the available throughput during peer-to-peer (P2P) communications. Yet, this has not been tested and verified in real-time indoor scenario, which has left much room for the investigation regarding the influences of localization accuracy on effective data signal strength.

**Table 1.** Comparisons of Relevant Literature with Our Work.

| Ref. | Direction | Mutual Communication Error Correction | 3-D Theory | 60 GHz | Field Test |
|---|---|---|---|---|---|
| [1] | × | × | √ | √ | × |
| [2] | √ | × | × | × | × |
| [4] | √ | × | √ | × | × |
| Proposed | √ | √ | √ | √ | √ |

While many of the afore-mentioned works have provided schemes and simulation analysis for mmWave communications, few have been made realized in practical environment. These have brought about the need for a real-world design and implementation of P2P communications under an appropriate combination of mmWave utilization and sensing-aided localization. The comparisons of relevant literature with our work are summarized in Table I concerning major contributions.

**1.2 Motivation and Challenge**

Motivated by these existing works, this paper selects three MikroTik wAP 60G AP routers as an access point (AP) and two clients in support for 60 GHz mutual communications, one Retina 4SN for the 60 GHz sensing technology, and two MPU9250 GY-9250 devices for IMU data collection and transmission. Two Raspberry Pi 3 (RPI) are employed for upper-layer control over information and data exchange among routers. Due to the inherent characteristics of mmWave and practical environments, there are three main challenges that need to be overcome.

(1) Scheduling. Joint-sensing-and-communication (JSAC) requires real time updates on both client locations and mutual communications, which take a toll on hardware and software implementations.
(2) Directionality. During dynamic location changing of clients, the beam directions need real-time calibration in order to beamform to AP to get orientation at each client and accurately beamform between clients to realize inter-communications.
(3) Noise elimination. Radar suffers from its blind estimation nature. The background noise and unrelated objects should be filtered from the desired objects to reduce target tracking inaccuracy.

**1.3 Approach and Contribution**

This paper investigates and proposes a comprehensive scheme of sensing-assisted P2P mmWave communications for dynamic client with real-field implementation. The main contributions are summarized as follows.

(1) A novel system that enables P2P mmWave communications with dynamic clients and sensing technologies is designed and deployed under the practical indoor environment.
(2) The combination of two algorithms for real-time JCAS, as well as P2P beamforming of mobile clients with appreciable accuracy, are studied. Specifically, for the tracking of clients, the density-based spatial clustering of applications with noise (DBSCAN) is incorporated for the accuracy of algorithms.
(3) The mmWave communication system is implemented in lab environment and tested for its performance with concrete results. It is revealed from the results that with mutual sensing and proposed algorithm, the communication performance of the system can be greatly improved.

## 2 Algorithm Design

A system that operates on a frame-to-frame basis is employed. During each frame, the radar module runs for $n$ seconds. We extract the data corresponding to second to last time-instance for guaranteed completeness and real-time of radar data. In the same frame, the latest IMU data is obtained from two clients and fed into the algorithm.

In the next stage, each frame of the algorithm runs around the extracted data. The radar data is passed through a tracking algorithm that calculates all clusters in view and their corresponding velocity. IMU data is used to correct and offset clients' true velocities. The cluster update algorithm is used for continuous tracking and update of clusters, after point clouds are collected with DBSCAN [12]. The reason we employ DBSCAN algorithm is that it is a density-based clustering method that has been widely used throughout year. It can locate the clusters of arbitrary shape and size and distinguish required objects from environment noise or unrelated objects [13]. Two key parameters for DBSCAN algorithm is: eps and Minpts, which are chosen based on the clients' real-time velocities and radar measurements rate.

The clients are identified by comparing cluster velocity with corrected IMU velocity and the recursive Kalman Filtering is leveraged to smooth the trajectory of moving clients to ensure the realization of client mobility. With the obtained location and orientation of mobile clients, the beamforming angle can be adjusted automatically in real-time to achieve directional communication among clients.

## 2.1 Cluster Localization

**Cluster Filtering.** The radar data also uses DBSCAN - returns a collection of point clouds and DBSCAN. The empirically-selected hyper-parameters, $Eps = 0.3$, $N = 100$, which are applied to identify all the clusters and filter out noise clusters. Then, a second layer of filtering is applied whereby all clusters with strictly zero doppler are regarded as background objects and filtered out, with the key observation being that even a stationary client has non-zero points.

**Cluster Update.** All remaining clusters are represented by corresponding core points calculated as centroids. During each frame, and beginning with the second frame, each cluster is mapped to its most likely cluster counterpart of the current frame. This is accomplished with the following Algorithm 1 *Cluster Update Algorithm*.

---
**Algorithm 1** Cluster Update Algorithm

**Input**: A list of clusters from current frame, $C$
**Input**: A list of clusters from previous frame, $C_p$
**Input**: Threshold = T
**Output**: A list of clusters from current frame $C_p$, $C_c$
**if** $|C_p == 0|$ **then**
    $C_c = C$
**else if** $|C_p| \leq |C|$ **then**
    $C_c = MLE(C_p)$
  **else**
    $C_c = MLE(C)$
  **end if**
  **if** $|C_c(\text{cluster A}) - C_p(\text{cluster A})| \geq T$ **then**
    Delete cluster A
  **end if**
**end if**

In the algorithm above, the Maximum Likelihood Estimator (MLE) is calculated as follows

$$C_c = \underset{C_c \in \mathbb{C}^{\min(|C_c|, |C_p|)}}{\arg\min} \|C_c - C_p\|, \quad (1)$$

where $|C|$ represents the cardinality of the set, and $\mathbb{C}$ represents the space formed from current list of cluster $|C|$ before matching. After calculating MLE in (1), the threshold in above algorithm is determined in the following way. Assume the client follows a random Gaussian process, in which situation we model the velocity as a Gaussian random variable with $v \in \mathcal{N}(v_m, v_r)$. The threshold is picked with certain percentage confidence as several standard deviations from the mean value of velocity.

We employ a very principled approach to velocity tracking. After IMU readings are calibrated, the velocity is calculated as the integral of acceleration over frame duration

$$V_c = V_p + \frac{A_c + A_p}{2} t \quad (2)$$

where $V_c, V_p, A_c, A_p$ are velocity and acceleration from the current and previous frame respectively, and $t$ is the frame time. The integration in (2) is discretized by choosing the average acceleration between current and previous frame IMU readings. The orientation of the client's router is tracked with the 6D Madgwick filter, after which the velocity is corrected by the orientation to convert IMU frame velocity to global frame velocity.

**2.2 Client Identification**

In order to identify the clients, the main idea is to find, with maximum likelihood, the corresponding clusters in $C_c$ that are clients, by velocity comparison. This is illustrated in the following Algorithm 2.

---
**Algorithm 2** Client Identification Algorithm

**Input**: A list of velocity from current frame clusters, $V_{\text{cluster}}$

**Input**: A list of Velocity of two clients, $V_{\text{client}}$

**Output**: A list of clusters corresponding to two clients

**if** Frame<2 or Frame>5 or No error, **then**
    Pass
**else** $\hat{V} = \underset{\hat{V} \in \mathbb{V}^2_{\text{cluster}}}{\arg\min} \|\hat{V} - V_{\text{client}}\|$
    **For** $i$ in $\hat{V}$
        Extract cluster label
    **end if**
**end if**

---

Note that the client identification is only called from frames 2 and 5, and whenever the cluster readings produce an error, either because the client moves outside the radar viewer, or it is non-line-of-sight (NLoS). Only on the second frame can the velocity lists be populated for identification to work. In addition, in order to avoid accumulated velocity drift from IMU, clients are not being continually identified after frame 5 whose tracking relies on *Algorithm 1 Cluster Update Algorithm*.

We smooth the trajectory of clients with a Kalman Filter (KF). There are two key observations. Firstly, during each frame, Kalman Filter extracts the measurement coordinate of clients by using client labels determined in client identification. This could err, when cluster update post frame 5 maps client cluster onto noise points, i.e. when client velocity is high, in which case *Cluster Update Algorithm* is run again. Secondly, the KF is discretized by having the average frame time as the interval. It is constructed based on the assumption that the clients follow Newtonian physics where the force exerted upon each client follows a Gaussian random variable.

After the filtered positions of both clients and their respective orientations are stored, the beamforming direction (angle) is calculated, which is necessary for peer-to-peer communication. The general condition is such that the clients must be in the beamspace of each other, i.e. a 180-degree span region centering the orientation of each client. After the beam angles are calculated with simple geometry, the Tx beamforming sector can be set on Mikrotik routers.

## 3    System Implementation

### 3.1    Network Infrastructure

The networking aspect of the system is implemented by leveraging a multithreaded server that continuously retrieves the latest IMU data from each client. The client-to-server communication link is realized by leveraging the UDP networking protocol in which each client is bound to a unique port on the server. During each frame, a thread-safe operation transfers the data from the individual client threads to a local object whose values can be read and matched with the most recent radar data.

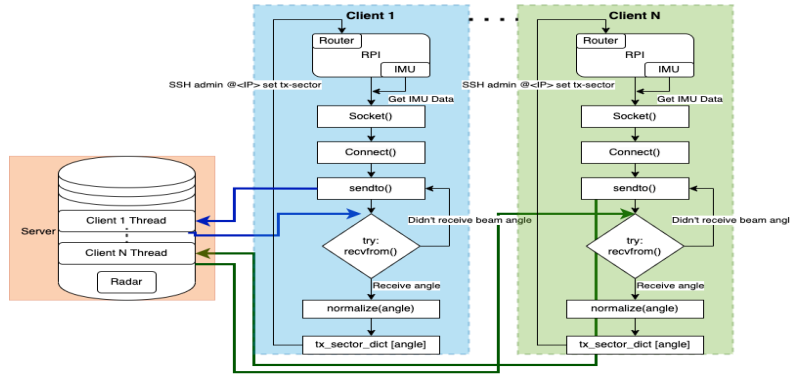

**Fig. 1.** IMU client back end. This flow chart describes the overall logic of the client-server communication link.

The design decision is made to minimize latency between the received data from the IMUs and the detection data from the radars. This creates a more consistent basis upon which these nearly simultaneous data points can be analyzed by the algorithm mentioned in the previous section.

Upon localization of the clients by the algorithm and beam angle calculation, the calculated angle between both clients is then communicated back to each client respectively. Once this angle feedback process is accomplished, each client will determine which antenna must be used to beamform and normalize the angle according to the best antenna choice as mentioned in Section 2.6 and retrieves the tx-sector value that corresponds with the calculated angle. This allows the RPI to *Secure Shell (SSH)* onto the router's admin user and set the tx-sector value accordingly.

### 3.2 Hardware Setting

The general hardware setting of the whole system is composed of three MikroTik 60G routers: one single antenna router as server, two three-antenna routers as clients, as well as RPI and a RETINA 4SN 60GHz Radar are shown in Fig. 2. Note that although two MikroTik 60G clients are set stationary as shown in Fig. 2, in the practical field test they are each hold by different testers and kept moving. Their velocities are thus always not equal to zero, with non-zero Doppler shift and changing 3D coordinates. Compared with stationary objects which are seen as ambient noise, the clients can be distinguished easily from them using previously-mentioned Kalman Filtering.

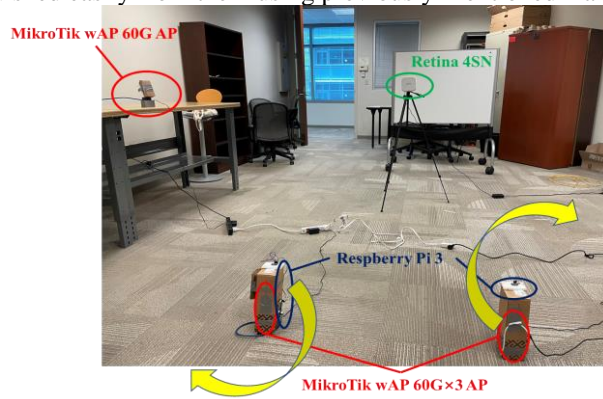

**Fig. 2.** General setting of related devices.

This system is set up with a Linux Desktop running Ubuntu 11.04 LTS, used to launch and run the algorithm. Each router can direct its beam to one of the many tx-sectors, which is a set of non-overlapping spatial region in its 60 by 30 radiation space. The clients are each set up with an RPI running the latest version of Raspbian OS. These are each wired to a three-antenna MikroTik 60G router and an MPU9265 IMU. The routers are enclosed in a cardboard box with additional cardboard strapped over the antenna. Then, some forms of manual attenuation can be achieved as there is no configurability over transmit power for the outdoor routers.

For the purposes of beamforming and data collection, the three-antenna routers are set to only use their middle antenna so that they would behave as a single antenna router. This is because the client routers are not capable of achieving direct communication with each other without going through the server, due to their configurations. Our results are accomplished by using our algorithm to localize the AP router which is a single-antenna router and a client router which is set to single-antenna mode. The angles are calculated accordingly and beam angles are set respectively.

This system is accomplished in an LoS environment between the radar and clients. The area is well-lit; however, lighting didn't have any implications on the radar or the algorithm's ability to track the clients. Additionally, both clients were initially still while the system stabilized and then were mobile for the remainder of the duration of tracking and beam angle calculation. The testing area is surrounded by desks as potential interference and the mobility of clients is constrained due to limited lab space.

## 4 Experimental Results and Discussion

### 4.1 Tracking Accuracy Performance

To evaluate the performance of the tracking algorithm, we follow a ground truth paths in Fig. 3, which are two green lines shaped like letter "P" respectively. The algorithm dynamically tracks the clients as they traverse the predefined path. The accuracy is calculated using the root mean square (RMS) error metric, where the true coordinate is obtained by projecting the measured coordinate onto its nearest dimension axis on the path. The RMS averaged over successive tests is approximately 16.24 cm.

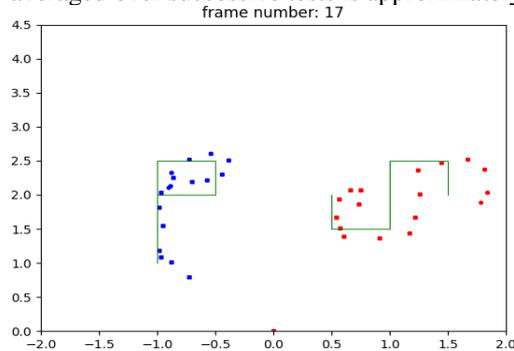

**Fig. 3.** The localization accuracy was evaluated with respect to a fixed ground truth pathing. The root mean square averaged over successive tests was approximately 16.24 cm.

The accuracy of tracking depends largely on IMU accuracy, as a suboptimal IMU could disrupt client identification in the presence of noise or other moving objects in the environment. When the environment is clear or has relatively low noise, our tracking algorithm is very reliable. In the test area, as there are several humans in the background, it is possible for the algorithm to "lock" onto other moving objects though the algorithm can correct itself after a number of frames. The limited frame update rate means the algorithm suffers from a high-mobility environment and clients. Finally, there are two additional sources of error that could be accounted for in a further extension of the research. Firstly, the radar point clouds are only the surface of a human body, causing the measured trajectory to drift towards radar. Secondly, eliminating elevation in core point tracking is susceptible to error if radar is not placed at an optimal height.

### 4.2 Beamforming Performance

Beamforming performance is evaluated with respect to each intersection in the ground truth paths in the previous Section 4.1. This is compared to routers' own ability to beam scan. The beam scan algorithm of routers is implemented such that multiple tx-sectors are "grouped together" and the routers slowly perform a search on each group and subgroups in between to determine which specific tx-sector value satisfies its threshold requirements for a stable connection. Once the tx-sector is determined, the received signal strength indicator (RSSI) and signal quality metrics are documented.

Similarly, for our algorithm, we allowed for the tx-sector to be calculated, set and then the signal quality and RSSI metrics measured. The signal quality metric is a transformation of signal-to-interference-plus-noise-ratio (SINR) created by MikroTik's RouterOS to provide an estimate for signal quality from 0 to 100. The results for both tests are shown in Table 2 and Table 3.

**Table 2.** Groundtruth RSSI *VS* Algorithm RSSI for each intersection.

| Intersection | Groundtruth | [4]-w/t sensing, Increase % | Proposed, Increase % |
|---|---|---|---|
| 1 | -58 | -56, 3.45% | -53, 8.62% |
| 2 | -53 | -53, 0.00% | -51, 3.77% |
| 3 | -53 | -52, 1.89% | -50, 5.66% |
| 4 | -55 | -54, 1.82% | -51, 7.27% |

**Table 3.** Groundtruth signal *VS* Algorithm Signal for each intersection. Signal is an arbitrary transformation of SINR generated by MikroTik's RouterOS.

| Intersection | Groundtruth | [4]-w/t sensing, Increase % | Proposed, Increase % |
|---|---|---|---|
| 1 | 85 | 92, 8.24% | 100, 17.65% |
| 2 | 90 | 92, 2.22% | 95, 5.56% |
| 3 | 70 | 82, 17.14% | 95, 35.71% |
| 4 | 75 | 86, 14.67% | 95, 26.67% |

The algorithm calculation of the beam angles has a cutting edge as opposed to the beam scan implementation which is already used by the routers and that without sensing in [4]. Additionally, it is important to note that because of the nature of beamscan, the RSSI and Signal Values continue to oscillate as the routers determine what the optimal tx-sector value is. These values cause the power to have a significant variance from 40 to 100 within the span of seconds. For our algorithm, this value stays more consistent and though it oscillates, it would stay from 90 to 100. This slight oscillation after algorithmically setting of the beam angle can likely be attributed to unavoidable noise and interference in the lab environment.

Furthermore, when looking at overall performance, we can see an average of 3.3% increase in RSSI values and roughly an average of 12.6% increase for the signal metric mentioned earlier. Specifically, when looking at intersections 1, 3, and 4 we notice a significant difference in the beam scan signal and the algorithm signal. This can likely be attributed to the routers not directly facing each other and being angled from each other. While the beamscan algorithm tries to seek the best tx-sector, the variance induced in signal quality is much more drastic than the signal variance from the sensing solution that we have created.

It is also valuable to note that the range on these routers is around 500 m and we were performing our experiments in a $25m^2$ space. This implies that even if the tx-sector is incorrectly set, the RSSI values will not see significant deviations because of the transmission power of the routers themselves, even after the manual attenuation that we implemented.

## 5    Conclusion and Future Applications

In this paper, we designed and implemented a peer-to-peer mmWave communication system assisted by mutual beamforming and radar sensing in an LOS indoor environment. With our proposed two algorithms, i.e. Cluster Update Algorithm and Client Identification Algorithm, we designed and completed a system using all its current capacity. The results show that the trajectories of target clients can be reliably tracked while simultaneously enabling beamforming capability. After further refinement and scale-up study, it is expected that our proposed system can be deployed in a larger environment, thus serving more clients simultaneously.

In the future, there are plausible applications based on our designed system. For example, when applied to smart home, high-speed communication between smart devices such as smart thermostats, security cameras, appliances, and lighting systems can be better enabled. In addition, for healthcare and industrial internet of things system, communication between devices and sensors will be facilitated and efficiency of monitoring systems, wearable devices, and other applications will be improved. The practical implementation, as well as extension, of our system will be of vital significance and convenience to our daily life.